**Magnetic Relaxometry of Methemoglobin by Widefield Nitrogen-Vacancy Microscopy**

Suvechhya Lamichhane,[1] Evelyn Carreto Guevara,[2] Ilja Fescenko,[3] Sy-Hwang Liou,[1] Rebecca Y. Lai,[2] and Abdelghani Laraoui[1,4,*]

[1]*Department of Physics and Astronomy and the Nebraska Center for Materials and Nanoscience, University of Nebraska-Lincoln, Lincoln, Nebraska 68588, USA*
[2]*Department of Chemistry, University of Nebraska-Lincoln, Lincoln, NE 68588, USA*
[3]*Laser Center, University of Latvia, Riga, LV-1004, Latvia*
[4]*Department of Mechanical & Materials Engineering, University of Nebraska-Lincoln, Lincoln, NE 68588, USA*
[*]*Corresponding author: alaraoui2@unl.edu*

## ABSTRACT

Hemoglobin (Hb) is a multifaceted protein, classified as a metalloprotein, chromoprotein, and globulin. It incorporates iron, which plays a crucial role in transporting oxygen within red blood cells. Hb functions by carrying oxygen from the respiratory organs to diverse tissues in the body, where it releases oxygen to fuel aerobic respiration, thus supporting the organism's metabolic processes. Hb can exist in several forms, primarily distinguished by the oxidation state of the iron in the heme group, including Methemoglobin (MetHb). Measuring the concentration of MetHb is crucial because it cannot transport oxygen, hence higher concentration of MetHb in the blood causes methemoglobinemia. Here, we use optically detected magnetic relaxometry of paramagnetic iron spins in MetHb drop-casted onto nanostructured diamond doped with shallow high density nitrogen vacancy (NV) spin qubits. We modify the MetHb concentration in the range of $6 \times 10^6 - 1.8 \times 10^7$ adsorbed $Fe^{+3}$ spins per $\mu m^2$ and observe an increase of the NV relaxation rate $\Gamma_1$ (= $1/T_1$, $T_1$ is NV spin lattice relaxation time) up to $2 \times 10^3$ $s^{-1}$. NV magnetic relaxometry of MetHb in phosphate-buffered saline solution shows a similar effect with an increase of $\Gamma_1$ to $6.7 \times 10^3$ $s^{-1}$ upon increasing the MetHb concentration to 100 μM. The increase of NV $\Gamma_1$ is explained by the increased spin noise coming from the $Fe^{+3}$ spins present in MetHb proteins. This study presents an additional usage of NV quantum sensors to detect paramagnetic centers of biomolecules at volumes below 100 picoliter.

Hemoglobin (Hb) is a crucial protein found in red blood cells (RBCs) responsible for transporting oxygen ($O_2$) from the lungs to tissues throughout the body,[1] ensuring cellular respiration and metabolic processes. This globular protein comprises four subunits, each containing a heme group that binds to $O_2$ molecules. The binding and release of $O_2$ by Hb are finely regulated to match the $O_2$ demands of tissues, a process influenced by factors such as partial pressure of $O_2$, pH, and temperature.[2] It has a molecular weight of 64.5 kDa.[3] However, alterations in the structure or function of Hb can lead to disruptions in $O_2$ transport and delivery, potentially compromising physiological processes. One such alteration occurs with the formation of Methemoglobin (MetHb), a derivative of hemoglobin wherein the iron (Fe) within the heme group undergoes oxidation from the ferrous ($Fe^{2+}$) to the ferric ($Fe^{3+}$) state.[4] Unlike normal Hb, MetHb is unable to bind $O_2$ reversibly, resulting in impaired $O_2$ delivery to tissues.[5] The accumulation of MetHb can lead to a condition known as methemoglobinemia, which is characterized by tissue hypoxia and cyanosis, among other symptoms.[6] Hence, the concentration levels of Hb and MetHb in blood serve as vital indicators or biomarkers for a wide array of health conditions and physiological states.

Various techniques have been employed to detect Hb, including electron paramagnetic resonance (EPR) spectroscopy [7,8] and nuclear paramagnetic resonance (NMR) spectroscopy.[9,10] However, these methods require larger quantities of Hb (in powder or liquid form) to produce detectable signals at sub 100 pL volumes. While there have emerged several techniques (*e.g.*, mass spectrometry,[11,12] Raman spectroscopy,[13,14] and fluorescence microscopy[15,16]) to measure small Hb concentrations, it remains difficult to non-destructively monitor in real time their evolving energetic state and molecular composition, where signals that are weaker, more localized and/or biologically specific. Recently, an alternative technique has emerged for measuring magnetic/electrical fields and temperature at the nanometer scale based on optical detection of quantum states of light and defect spin qubits in a non-inductive method that circumvents the challenges of NMR/EPR spectrometers. The platform is based on optical detection of electron spin resonances of nitrogen vacancy (NVs) centers in diamond.

The NV center is a spin-1 defect with electron spin properties that can be addressed optically,[17–20] could exhibit millisecond quantum coherence at room temperature,[21] making it ideal for quantum sensing, [21–24] nanoscale magnetometry,[25–29] and biosensing.[30–33] There have been several studies of NV magnetometry, including the detection of single proteins,[34] nuclear spin ensembles from nanoscale volumes,[35–38] weak paramagnetic individual hemozoin biocrystals,[39] and $[Fe(Htrz)_2(trz)](BF_4)$ spin crossover molecules.[40] Other works have been performed by monitoring the NV spin-lattice $T_1$ relaxation time in the presence of paramagnetic ions. $T_1$ relaxometry has proven effective in detecting ions such as $Gd^{+3}$ [41] and $Cu^{+2}$,[42] $Fe^{+3}$ within ferritin proteins[43,44] and Cyt-C proteins.[45] NV $T_1$ relaxometry was used recently in nanodiamonds to detect Hb in free radicals[46,47] where the relaxation rate in rat's blood increases in deoxyhemoglobin as compared with oxyhemoglobin ($S = 0$).[48] For MetHb the spin state is $S = 5/2$,[49] thus more reduction in the $T_1$ relaxation is expected due to the high magnetic moment and therefore strong dipolar coupling with NV spins. Magnetic imaging utilizing NVs has recently emerged as a promising approach for achieving nanoscale resolution.[25–27] In this study, the NV relaxation rate $\Gamma_1$ ($= 1/T_1$) was measured on MetHb proteins in both solution and dried forms by using widefield NV microscopy in conjunction with X-ray photoelectron spectroscopy (XPS) and atomic force microscopy (AFM) to analyze MetHb proteins and assess their dimensions and characteristics.

The human Hb used in this work was purchased from Sigma Aldrich, which is a lyophilized powder, predominantly existing in the oxidized state known as MetHb, where the iron is in the $Fe^{+3}$ oxidation state. The molecular structure of hemoglobin is depicted in **Fig. 1(a)** and **Fig. 1(b)**. The pH of different hemoglobin solutions was varied depending on the solvent used and the concentration of hemoglobin. In phosphate-buffered saline (PBS) at pH 7.4 without protein, hemoglobin solutions showed pH values ranging from 7.54 to 7.60, whereas in deionized (DI) water (pH = 6.5) addition of Hb gives the pH values ranged from approximately 6.44 to 6.79.[50–52] Hemoglobin exhibited the most stability at physiological pH (~ 7.4), however MetHb is found to be stable within the pH range of 6.5 to 8.[53] The concentration of MetHb was varied from 30 μM to 100 μM in a PBS buffer on a diamond chip. MetHb proteins were diluted in DI water (concentration of 2 μM), drop-casted on the diamond surface. AFM was used to measure the size and height of MetHb nanoclusters as shown in **Fig. 1(c)**. The MetHb size analysis shows a variation of the nanocluster's diameter from 50 to 200 nm, whereas the height distribution shows an average height of ~ 10 ± 6 nm (see **Fig. 1 (d)**). The diameter of MetHb in its physiological state is reported to be 5 nm that varies depending on the pH values.[54] However, it has the ability to crystallize without undergoing denaturation or breaking down into its individual subunits.[55] The detection of MetHb proteins was realized via magnetic dipole-dipole interaction between the fluctuating $Fe^{+3}$ in MetHb and NV spins,[56] as explained below together with NV magnetic relaxometry technique.

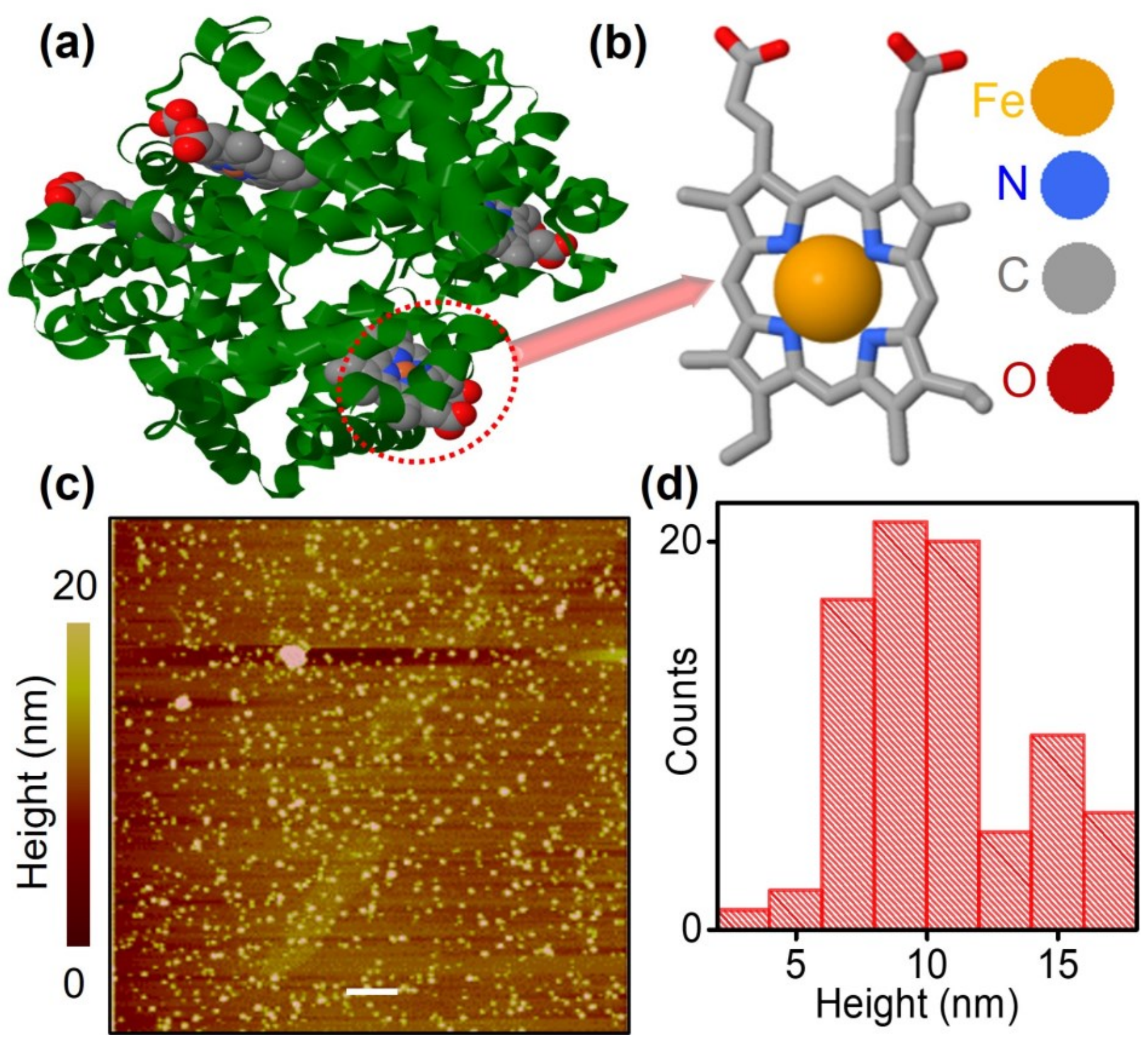

**FIG. 1.** (a) Ribbon structure of Hb complex derived from reference 1. (b) Molecular structure showing the heme center. (c) AFM image of the MetHb nanoclusters. The scale bar in (c) is 1 μm. (d) AFM height distribution of the MetHb nanoclusters (concentration of 2 μM) drop-cased on top of the diamond substrate. The mean height of MetHb nanoclusters is ~ 10 ± 6 nm.

To confirm the absence of impurities in MetHb, XPS measurements were performed using Thermo Scientific Al K-Alpha XPS system, operating under ultrahigh vacuum conditions with a pressure of $1 \times 10^{-9}$ mbar. Data acquisition and analysis were performed utilizing the Avantage™ software package. To mitigate charge effects during measurement, a combination of electron and argon ion flood guns was employed, maintaining an argon pressure in the chamber ranging from $2 \times 10^{-8}$ to $4 \times 10^{-8}$ mbar. The X-ray beam size was 400 μm, and high-energy resolution spectra were recorded with a pass energy of 50 eV, utilizing a step size of 0.1 eV and a dwell time of 50 ms. The number of averaged sweeps for each element was adjusted to optimize the signal-to-noise ratio, typically ranging from 20 to 100 sweeps. **Fig. 2(a)** shows the measured high-resolution XPS spectra of Fe 3p fitted with combined Lorentzian and Gaussian functions. The spectra exhibit a distinct peak at a binding energy of 55.6 eV, indicative of the Fe oxidation state being $Fe^{+3}$.[45] The XPS results obtained for C 1s, N 1s, and O 2p are plotted in **Figures 2(b)**, **2(c)**, and **2(d)**, respectively, confirm the absence of impurities in the studied MetHb proteins.

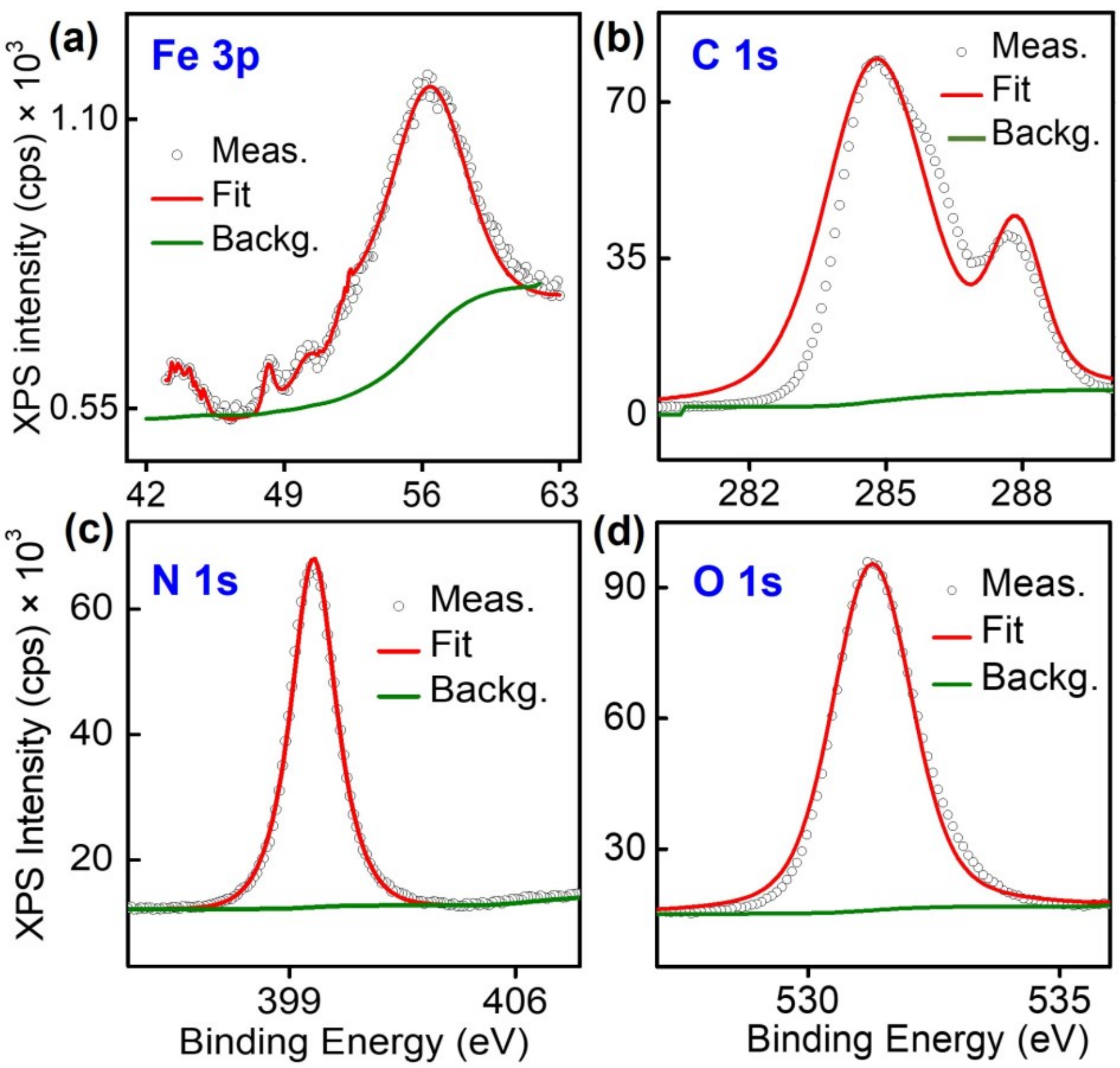

**FIG. 2.** High resolution XPS peak of MetHb with the background for (a) Fe 3p, (b) C 1s, (c) N 1s, and (d) O1s. Peak deconvolution is performed for each of the spectra. The background correction was performed using the smart background subtraction feature within the Avantage™ package. Spectral analysis involved peak deconvolution, which utilized a combination of Gaussian and Lorentzian functions. The residue background signal is highlighted by the solid green line.

The negatively charged NV center consists of a substitutional nitrogen atom adjacent to a vacancy site (**Fig. 3(b)**), possessing a spin triplet in its ground state with a zero-field splitting $D$ = 2.87 GHz between states $m_S$ = 0 and $m_S$ = ±1, **Fig. 3(c)**.[17,20] When exposed to green laser illumination (532 nm), spin-conserving excitation occurs, transitioning the NV center to an excited triplet state, subsequently emitting far-red photoluminescence (650 ─ 750 nm). The spin-dependent rate of intersystem crossing leads to the optical polarization of NVs into the $m_S$ = 0 state.[17] Microwave (MW) excitation allows spin transitions from $m_S$ = 0 to $m_S$ = ±1 and the applied magnetic field $B_{app}$ breaks the degeneracy of the $m_S$ = ±1 due to Zeeman splitting, leading to a pair of spin transitions ($m_S$ = 0 to $m_S$ = +1 and $m_S$ = 0 to $m_S$ = -1) that can be measured via optically detected magnetic resonance (ODMR) spectroscopy. The energy level of the center are such that NV centers in the $m_S$ = ±1 levels have decreased fluorescence versus the electrons in the $m_S$ = 0 level.[17] Thus, as the MW field is swept across the resonance spectra, the NV centers exhibit characteristic ODMR dips at the resonance frequency with a narrow (< 0.1 mT) linewidth, **Fig. 3(e)**. Since this resonance position depends linearly on the magnetic field component along the NV symmetry axis, the resonance position provides a highly accurate reading of the amplitude of the applied magnetic field.[17]

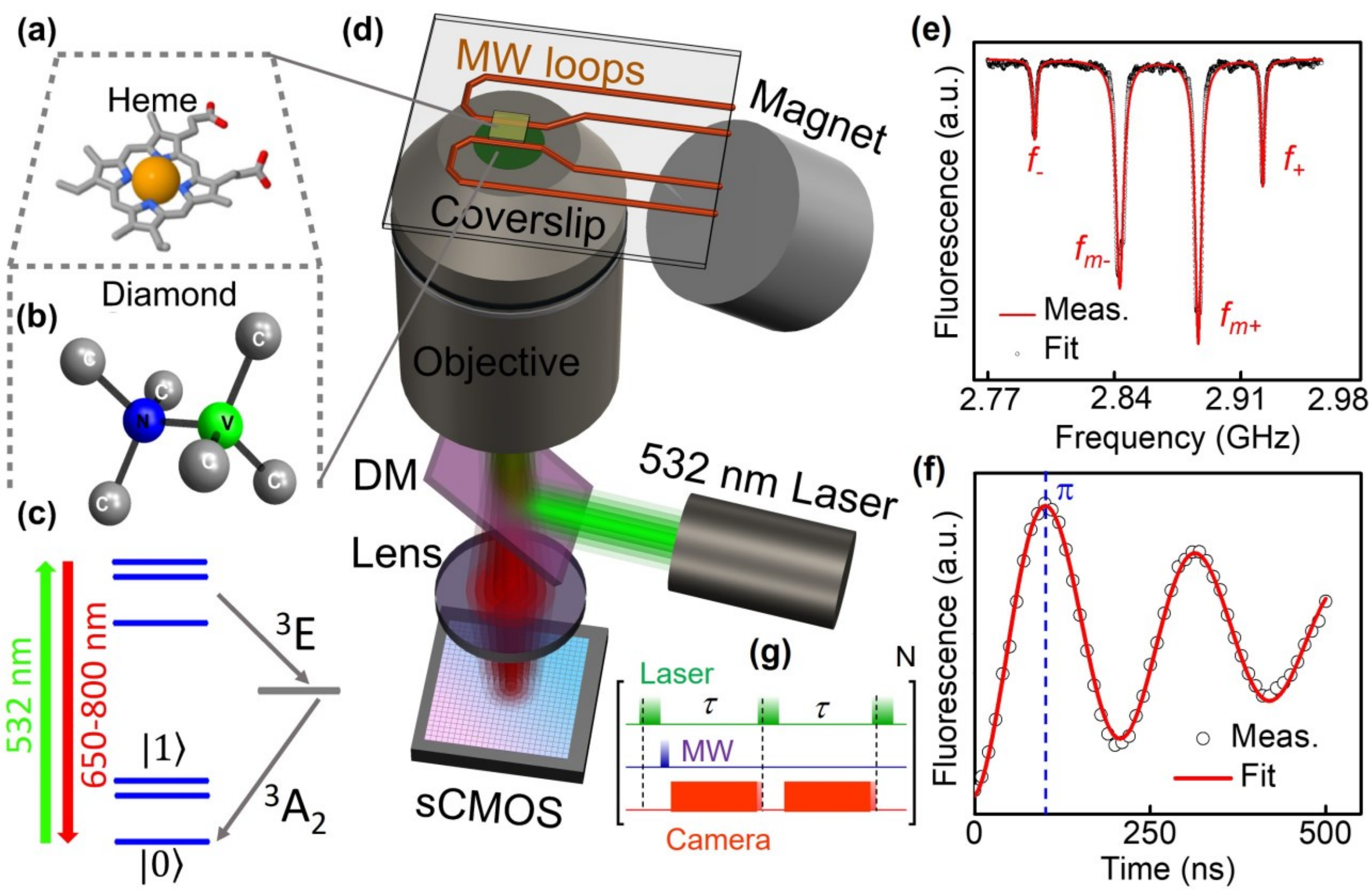

**FIG. 3.** (a) Molecular structure of the heme center of Hb. (b) A schematic of the NV center inside the diamond lattice (nitrogen: green atom, yellow: vacancy). (c) A schematic of the energy levels of the NV center ground ($^3A_2$) and excited ($^3E$) states with intermediate metastable state. (d) A schematic of the widefield NV microscope used for relaxometry of MetHb proteins. A magnetic field $B_{app}$ is applied along the [111] direction in (100) diamond. DM is a dichroic mirror. (e) NV ODMR spectrum at an applied field $B_{app}$ = 3.2 mT on a bare diamond aligned along the NV axis. (f) Rabi oscillation of $m_S$ = 0 state to $m_S$ = -1 spin transition at a MW frequency of 2.78 GHz and power of 200 mW. (g) A schematic of $T_1$ relaxometry technique pulse sequence. A laser pulse is used first to initialize the NV spins in $m_S$ = 0, then a MW $\pi$ pulse is applied to flip the NV spin to $m_S$ = -1, and finally a laser pulse is applied to read out the NV fluorescence. The sequence with $\pi$ pulse is subtracted from another sequence without $\pi$ pulse to get *rid of the common noise signal*. This sequence is repeated N times to accumulate the NV $T_1$ signal.

In **Fig. 3(d)**, we depict the widefield NV microscope[26,40] used for mapping MetHb proteins (**Fig. 3(a)**) drop-casted onto the diamond doped with NV spins. A 532-nm laser (power = 180 mW) is used to excite NVs over an area of 36 × 36 μm$^2$ and the NV fluorescence (650 ─ 750 nm) is mapped onto a sCMOS camera. Detailed information regarding the experimental setup is provided in our previous study on cytochrome c protiens.[45] A 3 mm × 2.5 mm × 0.05 mm electronic grade (100) diamond was cut and polished followed by $^{15}N^+$ implantation (4 keV), high temperature (1123 K) annealing under high vacuum (2.7 × 10$^{-6}$ mbar), and cleaning in a boiling tri-acid mixture. This procedure led a creation of a thin NV sensing layer below ~6 nm from the surface.[37,39,57] Then, the NV doped diamond substrate with MetHb protein was placed on top of the coverslip patterned with MW striplines as depicted in **Fig. 3(d)**. Upon the application of an applied magnetic field $B_{app}$ of 3.2 mT, four ODMR peaks appear in the spectrum (**Fig. 3(e)**), corresponding to NV sub-ensembles with different symmetry axes: $f_-$, $f_+$ for NVs aligned along the [111] direction and $f_{m-}$, $f_{m+}$ for NVs merged along the other direction.[40] After ODMR measurements, Rabi oscillations were measured (**Fig. 3(f)**) of t= 0 to $m_S$ = -1 spins aligned along $B_{app}$ (3.2 mT) to find the $\pi$-pulse width, a time needed to bring NV population from $m_S$ = 0 to $m_S$ = -1.

The relaxation measurements pulse sequence is depicted in **Fig. 3(g)**. The measurements process starts first by optically polarizing the NV center into $m_S$ = 0 state. A MW $\pi$ pulse is then applied in the first measurement allowing the NV $m_S$ = -1 spin state to evolve in the absence of the external influence of laser in dark environment for a time $\tau$. In the second reference measurement without the MW $\pi$ pulse the relaxation occurs from the state $m_S$ = 0. This process is repeated with a readout of the fluorescence contrast after each of two consequent measurements. The readout is done by a sCMOS camera with a frame size of 550 pixels × 550 pixels equivalent to a field of view of 36 μm × 36 μm at an exposure time of 4 ms. The camera exposure was overlapped with the laser pulse such that only the first microsecond of the laser induced fluorescence pulse was mapped on the camera, **Fig 3(g)**. In the simplest experiments where only the mean relaxation is measured, a fast avalanche photon detector (APD, Thorlabs APD410A) was used, connected to a Yokogawa oscilloscope (DL9041L)[40] reflecting the NV fluorescence by a flip mirror, which placed just before the sCMOS camera, and focused with a lens (focal length of 30 mm). This allows shorter readout times, thus increasing the signal-to-noise ratio. Pairs of the contrast measurements are collected at various $\tau$ times and averaged over many cycles (N = $10^4$) at each $\tau$ point. The final relaxation (fluorescence intensity *vs* $\tau$) curve is the subtraction of signal with and without MW $\pi$ pulse to remove other (*e.g.*, thermal) effects not related to magnetic noise.[58,59]

Finally, the measured relaxation curve was fitted with an exponential decay function to obtain $T_1$ values. To assess the spin noise produced by $Fe^{+3}$ spins in MetHb proteins, the NV relaxation rate $\Gamma_1 = 1/T_1$ is more convenient representation of NV relaxometry experiments due to the linearity.[45] The $Fe^{+3}$ spins in MetHb nanoclusters generate a fluctuating magnetic field that interacts with NV spins via dipolar magnetic interactions and increases its relaxation rate $\Gamma_1$, which depends on the dipolar interaction strength between $Fe^{+3}$ and NV spins and the fluctuation rate of the $Fe^{+3}$ spins as follows:[41]

$$\Gamma_1 = \frac{2\langle B^2\rangle f_t}{f_t^2 + D^2}, \qquad \textbf{(Eq. 1)}$$

where, $f_t$ is the fluctuation rate of $Fe^{+3}$ spins, $\langle B^2\rangle$ is the mean dipolar magnetic coupling strength between the NV spins and the $Fe^{+3}$ spins, and $D$ is zero field splitting of ground state. To estimate the depth of the NV layers $d$ from the diamond surface, we measured $\Gamma_1$ in $CuSO_4$ solution of different concentrations and found $d$ of 5.5 nm (see reference 45 for further details).

To distinguish the effect of $Fe^{3+}$ spins within MetHb proteins from other paramagnetic impurities inside the diamond or at its surface,[60,61] the diamond substrate was nanostructured with a silicon nitride (SiN) grating (height is 50 nm and spacing distance of 4 μm). See reference 45 for further details on the nanofabrication of SiN grating and its properties.

Prior to measuring on MetHb diluted in a PBS buffer solution, NV relaxation spectroscopy on diamond was performed with ($\Gamma_{bar}$=1.1 × $10^3$ $s^{-1}$) and without PBS ($\Gamma_{buf}$ = 0.8 × $10^3$ $s^{-1}$) at different locations using APD connected to the oscilloscope for $T_1$ spectroscopy readout. A decrease of 0.3 × $10^3$ $s^{-1}$ in $\Gamma_1$ is obtained with the introduction of PBS buffer (**Fig. 4(a)**), which is attributed to the diamagnetic behavior of the water molecules within the PBS, which suppress the surface electron noise.[62] The variation of $\Gamma_1$, which is found to be ~ 20%, comes primarily from the inhomogeneous distribution of NVs within the diamond substrate,[45] which can modulate the charge state of the NV centers and the phonon bath surrounding them, respectively.[62] Next, NV $T_1$ relaxometry measurements on MetHb in a PBS solution were performed. The PBS buffer used had a concentration of 2 mM, with a pH of 7.60. $\Gamma_1$ is increased to $2.0 \times 10^3$ $s^{-1}$ after adding 30 μM MetHb, **Fig. 4(b)**. By increasing the concentration of MetHb in PBS to 50 μM, $\Gamma_1$ is further increased to $3.5 \times 10^3$ $s^{-1}$ (**Fig. 4(b)**), explained by the increased spin noise. **Eq.1** is used to fit the relaxation rate $\Gamma_1$ for the MetHb/PBS solution of various concentrations as shown by the solid red line in **Fig. 4(c)**. For estimating the external relaxation rate, the relaxation rate of the PBS buffer $\Gamma_{buf}$ was subtracted. In our calculation, the variance $\langle B^2 \rangle$ was derived from only the dipolar coupling between NV spins and $Fe^{+3}$ spins present in MetHb as a variable parameter. This dipolar coupling field strength amounts to 0.27 mT for a concentration of 100 μM which gives a relaxation rate of $6.7 \times 10^3$ $s^{-1}$. The fluctuation rate of Hb is noted as 0.5 GHz.[63]

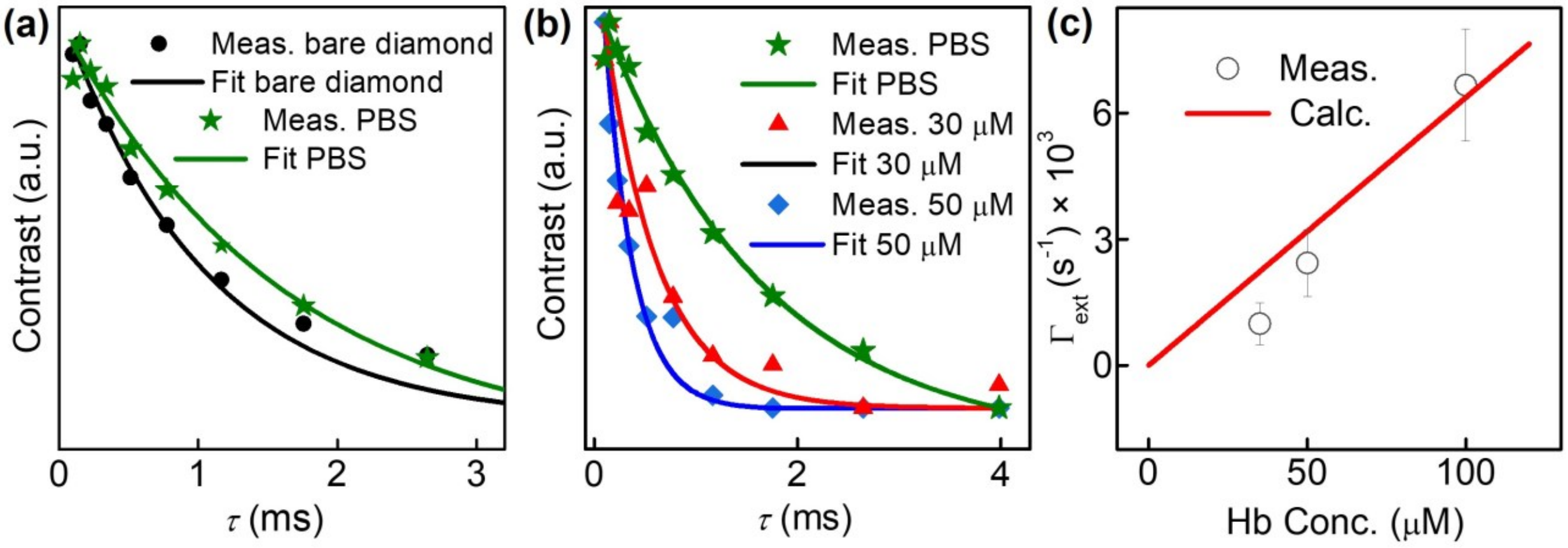

**FIG. 4.** (a) Measured $\Gamma_1$ relaxation of a bare diamond (filled stats) and a diamond with the PBS buffer solution (open squares), fitted (solid lines) with an exponential decay function to extract $\Gamma_1$. The measurements were performed using the APD and oscilloscope. (b) Measured NV fluorescence contrast as function of *t* for a diamond with PBS (filled stars), MetHb solution with a density of 30 μM (filled triangles), and MetHb solution with a density of 50 μM (filled diamonds). The measurements were fitted with exponential decay functions (solid lines) to extract $\Gamma_1$ values. (c) Measured (open circles) and calculated (solid line) relaxation rate $\Gamma_1$ of NV spins as function of the concentration of MetHb in PSB.

In the second set of the experiments, the relaxation rate induced by dried MetHb nanoclusters was imaged by using the pulse sequence in **Fig. 3(g)**. MetHb diluted in DI water was drop-casted on the diamond substrate similar to **Fig. 1(c)**. **Figures 5(a)**, **5(b)**, and **5(c)** show $\Gamma_1$ images of diamond with SiN grating with MetHb proteins adsorbed onto the diamond surface with three $Fe^{+3}$ spin densities of $6 \times 10^6/\mu m^2$, $1.2 \times 10^7/\mu m^2$, and $1.8 \times 10^7/\mu m^2$ respectively. The region with SiN grating gives the intrinsic relaxation rate of $\Gamma_{int}$ of the bare diamond, *i.e.*, surface effects reduced.[45] **Fig. 5(d)** displays the horizontal line cut integrated with all pixels for the spin density $1.2 \times 10^7/\mu m^2$. The measured $\Gamma_{sig}$ of $2.0 \times 10^3$ $s^{-1}$ is the contribution from the spin noise due to $Fe^{+3}$ spins in MetHb and the surface impurities from the diamond lattice ($\Gamma_{sur} \approx 0.6 \times 10^3$ $s^{-1}$), whereas $\Gamma_{int}$ in the intrinsic diamond relaxation rate ($0.5 \times 10^3$ $s^{-1}$). The relaxation rate change from only MetHb nanoclusters is $\Gamma_{ext} \approx \Gamma_{sig} - \Gamma_{int} - \Gamma_{sur} = 0.9 \times 10^3$ $s^{-1}$. It is noteworthy that the SiN layer of the grating gets etched after repetitive cleanup process using piranha solution to eliminate any magnetic impurities. Consequently, a complete suppression of the surface diamond spin noise could not be achieved.

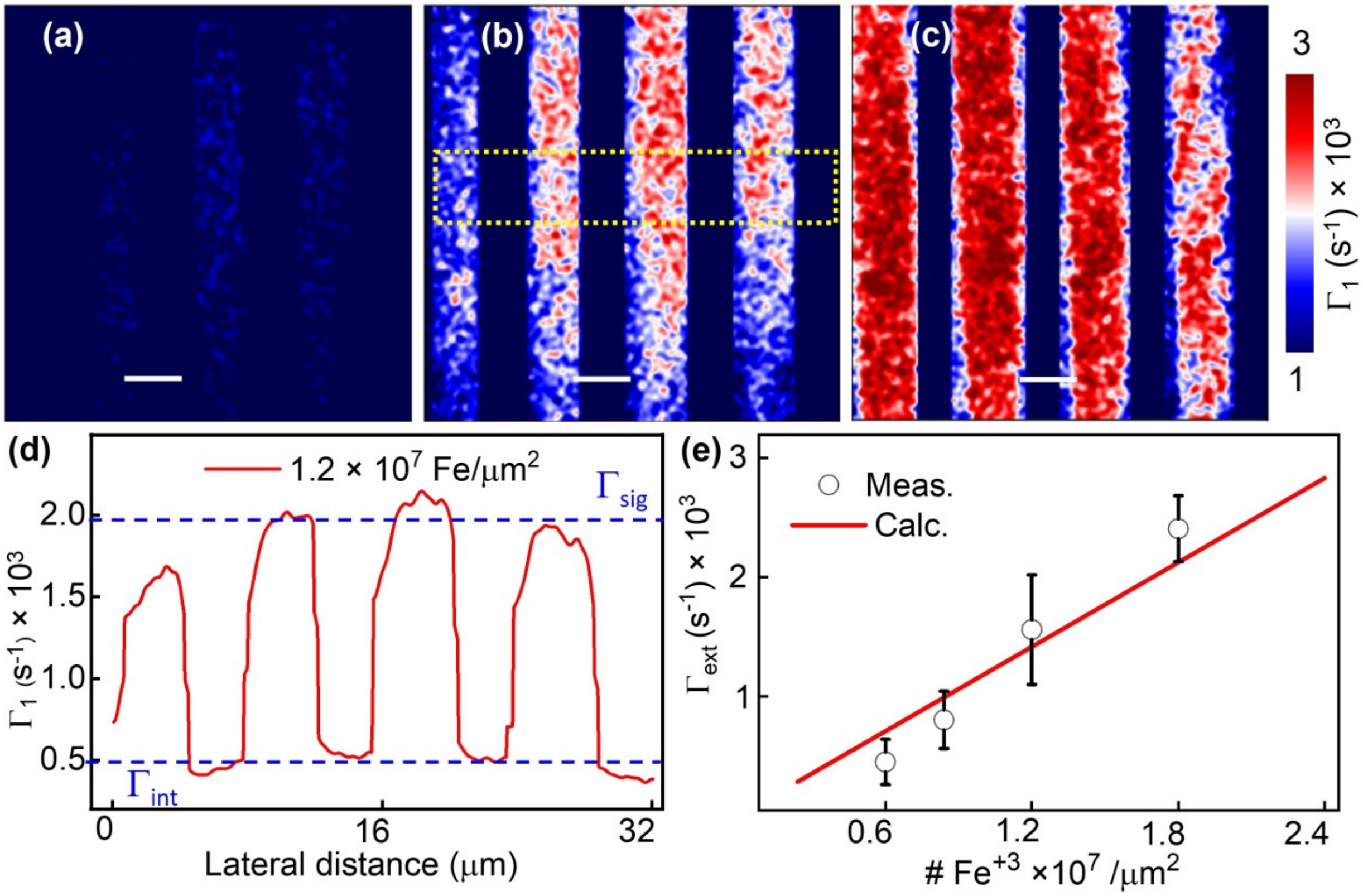

**FIG. 5.** $\Gamma_1$ images of MetHb nanoclusters drop-casted on diamond with $Fe^{+3}$ spins concentration of $6 \times 10^6/\mu m^2$ (a), $1 \times 10^7/\mu m^2$ (b), and $1.8 \times 10^7/\mu m^2$ (c), respectively. The scale bar in (a), (b), and (c) is 5 μm. (d) Corresponding extracted horizontal $\Gamma_1$ profile from (c). (e) Measured (open circles) and calculated (solid line) $\Gamma_{ext}$ as function of the density of $Fe^{+3}$ spins in Hb adsorbed on diamond.

**Fig. 5(e)** shows the $\Gamma_{ext}$ plotted against the spin density demonstrated by open circles. The theoretical dependence (solid line in **Fig. 5(e)**) of $\Gamma_{ext}$ *vs* the density of $Fe^{+3}$ adsorbed centers per 1 $\mu m^2$ was calculated by using **Eq. 1**. For theoretical estimation, the variance $\langle B^2 \rangle$ obtained from the dipolar coupling of NV spins and external spins was kept as a free parameter. The dipolar coupling field strength is 0.107 mT and the fluctuation rate of MetHb is 0.5 GHz for the spin density of $1.7 \times 10^7$ $Fe^{+3}$ adsorbed/ $\mu m^2$.

Previous investigations illustrated $\Gamma_1$ analysis using nanodiamonds in fresh rat blood, primarily consisting of oxygenated blood (> 96%).[48] Since hemoglobin in erythrocytes confines direct interaction with NV, $\Gamma_1$ is increased from $0.8 \times 10^3$ $s^{-1}$ to $1.1 \times 10^3$ $s^{-1}$ within the blood itself. Moreover, $\Gamma_1$ increase was observed with hemolysis due to heightened dipolar interaction between NV spins and hemoglobin following erythrocyte rupture to ~ $1.3 \times 10^3$ $s^{-1}$.[48] The later study directly measured the relaxation change in blood which contains other ions or diamagnetic substances which could not solely be the relaxation of hemoglobin. It is difficult to compare with our study where we directly measure on lipolyzed MetHb powder. Additionally, in our study, the purchased hemoglobin is oxidized to methemoglobin, consequently resulting in a significant increase in $\Gamma_1$. The minimum detectable $Fe^{+3}$ spin density in MetHb from our NV measurement is $0.6 \times 10^7$ /$\mu m^2$. To improve further the sensitivity, one can selectively functionalize the diamond surface to immobilize the proteins in a controlled manner, facilitating their molecular-level study, and using single NVs with high ODMR and $T_1$ contrasts.[64]

In summary, we demonstrated the detection of change in concentration of human MetHb in solution and dried nanoclusters (size: 50 – 300 nm, height: 5 – 15 nm) by using NV $\Gamma_1$ relaxometry. NV optical detected magnetic relaxometry allows the detection of spin-noise generated by $Fe^{+3}$ spins in MetHb via the increase of $\Gamma_1$ up to $6.7 \times 10^3$ $s^{-1}$ for 100 μM MetHb solution, corresponding to a concentration of ~ $6.23 \times 10^3$ molecules. The higher sensitivity to MetHb in solution may be attributed to the more uniform distribution of proteins within the solution. However, during drop casting and drying processes, the deposition of nanoclusters may not be uniform, leading to the formation of voids. Consequently, the addition of MetHb beyond the sensing layer may exceed the detection range for changes in $\Gamma_1$. By patterning the diamond with SiN grating, we adsorbed MetHb molecules with varying spin density and imaged the relaxation rate $\Gamma_1$ *vs* the spin density of adsorbed Fe spins in MetHb per 1 $\mu m^2$ area. Measurements of MetHb in dried and solution forms show similar effects with a slight effect of PBS buffer due to the diamagnetic behavior. The measured NV relaxation rates agree well with the calculated values deduced from a model of interacting $Fe^{+3}$ centers with NV spins in the diamond substrate with a depth of 5.5 nm.

Surface functionalization with enhanced NV sensitivity is crucial for the detection of sub-microscale hemoglobin volumes. Recent studies have shown that employing slow oxidative etching of implanted diamond followed by cleaning in a boiling tri-acid mixture can produce NVs as shallow as 2 nm from the surface, resulting in prolonged $T_1$ and $T_2$ times suitable for sub-2 $nm^3$ magnetic resonance imaging. Another approach to enhancing the coherence time of NV spins involves minimizing surface spins by annealing and cleaning the diamond in oxygen rich environment.[14] Consequently, positioning individual biomolecules within the 5-nm sensing range of a single NV center enables conducting EPR and NMR spectroscopy on individual-intact biomolecules. Integrating microfluidics[65,66] and environmental control[64,67] with enhanced sensitivity NV magnetometer may allow for trace analysis of MetHb in blood samples.

## ACKNOWLEDGMENTS

This material is based upon work supported by the NSF/EPSCoR RII Track-1: Emergent Quantum Materials and Technologies (EQUATE) Award OIA-2044049 and the NSF Award 2426522. I.F. acknowledges support from the Latvian Quantum Initiative under European Union Recovery and Resilience Facility Project No. 2.3.1.1.i.0/1/22/I/CFLA/001. The research was performed in part in the Nebraska Nanoscale Facility: National Nanotechnology Coordinated Infrastructure and the Nebraska Center for Materials and Nanoscience (and/or NERCF), which are supported by the NSF under Award No. ECCS: 2025298, and the Nebraska Research Initiative.

## AUTHOR DECLARATIONS

### Conflict of Interest

The authors have no conflicts to disclose.

## Author Contributions

**Suvechhya Lamichhane**: Investigation (lead), Formal analysis (Lead), Writing – original draft (lead); Writing – review & editing (equal). **Evelyn Carreto Guevara**: Data Curation (supporting); Writing – review & editing (equal). **Ilja Fescenko**: Validation (equal); Writing – review & editing (equal); Formal analysis (supporting). **Sy-Hwang Liou**: Validation (equal); Supervision (equal); Writing – review & editing (equal); Funding acquisition (equal). **Rebacca Lai**: Conceptualization (equal); Supervision (equal); Writing – review & editing (equal); Funding acquisition (lead). **Abdelghani Laraoui**: Conceptualization (lead); Supervision (lead); Writing – review & editing (equal); Funding acquisition (equal).

## DATA AVAILABILITY

The data that support the findings of this study are available from the corresponding author upon reasonable request.

## REFERENCES

[1] M.H. Ahmed, M.S. Ghatge, and M.K. Safo, "Hemoglobin: Structure, Function and Allostery," Subcell Biochem **94**, 345–382 (2020).

[2] J.W. Adamson, and C.A. Finch, "Hemoglobin function, oxygen affinity, and erythropoietin," Annu Rev Physiol **37**, 351–369 (1975).

[3] H.H. Billett, "Hemoglobin and Hematocrit," in *Clinical Methods: The History, Physical, and Laboratory Examinations*, edited by H.K. Walker, W.D. Hall, and J.W. Hurst, 3rd ed., (Butterworths, Boston, 1990).

[4] B.K. Biswal, and M. Vijayan, "Structures of human oxy- and deoxyhaemoglobin at different levels of humidity: variability in the T state," Acta Cryst D **58**, 1155–1161 (2002).

[5] S.S. Cho, Y.D. Park, J.H. Noh, K.O. Kang, H.J. Jun, and J.S. Yoon, "Anesthetic experience of methemoglobinemia detected during general anesthesia for gastrectomy of advanced gastric cancer -A case report-," Korean J Anesthesiol **59**(5), 340–343 (2010).

[6] A. Iolascon, P. Bianchi, I. Andolfo, R. Russo, W. Barcellini, E. Fermo, G. Toldi, S. Ghirardello, D. Rees, R. Van Wijk, A. Kattamis, P.G. Gallagher, N. Roy, A. Taher, R. Mohty, A. Kulozik, L. De Franceschi, A. Gambale, M. De Montalembert, G.L. Forni, C.L. Harteveld, J. Prchal, and SWG of red cell and iron of EHA and EuroBloodNet, "Recommendations for diagnosis and treatment of methemoglobinemia," American Journal of Hematology **96**(12), 1666–1678 (2021).

[7] M. Lores, C. Cabal, O. Nascimento, and A.M. Gennaro, "EPR study of the hemoglobin rotational correlation time and microviscosity during the polymerization of hemoglobin S," Appl. Magn. Reson. **30**(1), 121–128 (2006).

[8] J. Peisach, W.E. Blumberg, B.A. Wittenberg, J.B. Wittenberg, and L. Kampa, "HEMOGLOBIN A: AN ELECTRON PARAMAGNETIC RESONANCE STUDY OF THE EFFECTS OF INTERCHAIN CONTACTS ON THE HEME SYMMETRY OF HIGH-SPIN AND LOW-SPIN DERIVATIVES OF FERRIC ALPHA CHAINS*," Proc Natl Acad Sci U S A **63**, 934–939 (1969).

[9] B.K. Fetler, V. Simplaceanu, and C. Ho, "1H-NMR investigation of the oxygenation of hemoglobin in intact human red blood cells," Biophys J **68**(2), 681–693 (1995).

[10] R. Macdonald, B.J. Mahoney, K. Ellis-Guardiola, A. Maresso, and R.T. Clubb, "NMR experiments redefine the hemoglobin binding properties of bacterial NEAr-iron Transporter domains," Protein Science **28**(8), 1513–1523 (2019).

[11] R. Théberge, S. Dikler, C. Heckendorf, D.H.K. Chui, C.E. Costello, and M.E. McComb, "MALDI-ISD Mass Spectrometry Analysis of Hemoglobin Variants: a Top-Down Approach to the Characterization of Hemoglobinopathies," J Am Soc Mass Spectrom **26**(8), 1299–1310 (2015).

[12] J. Zhang, G.R. Malmirchegini, R.T. Clubb, and J.A. Loo, “Native Top-Down Mass Spectrometry for the Structural Characterization of Human Hemoglobin,” Eur J Mass Spectrom (Chichester, Eng) **21**(3), 221–231 (2015).
[13] B.R. Wood, K. Kochan, and K.M. Marzec, “Chapter 13 - Resonance Raman spectroscopy of hemoglobin in red blood cells,” in *Vibrational Spectroscopy in Protein Research*, edited by Y. Ozaki, M. Baranska, I.K. Lednev, and B.R. Wood, (Academic Press, 2020), pp. 375–414.
[14] B.L. Dwyer, L.V.H. Rodgers, E.K. Urbach, D. Bluvstein, S. Sangtawesin, H. Zhou, Y. Nassab, M. Fitzpatrick, Z. Yuan, K. De Greve, E.L. Peterson, H. Knowles, T. Sumarac, J.-P. Chou, A. Gali, V.V. Dobrovitski, M.D. Lukin, and N.P. de Leon, “Probing Spin Dynamics on Diamond Surfaces Using a Single Quantum Sensor,” PRX Quantum **3**(4), 040328 (2022).
[15] I. Saytashev, R. Glenn, G.A. Murashova, S. Osseiran, D. Spence, C.L. Evans, and M. Dantus, “Multiphoton excited hemoglobin fluorescence and third harmonic generation for non-invasive microscopy of stored blood,” Biomed Opt Express **7**(9), 3449–3460 (2016).
[16] W. Zheng, D. Li, Y. Zeng, Y. Luo, and J.Y. Qu, “Two-photon excited hemoglobin fluorescence,” Biomed Opt Express **2**(1), 71–79 (2010).
[17] M.W. Doherty, N.B. Manson, P. Delaney, F. Jelezko, J. Wrachtrup, and L.C.L. Hollenberg, “The nitrogen-vacancy colour centre in diamond,” Physics Reports **528**(1), 1–45 (2013).
[18] A. Laraoui, F. Dolde, C. Burk, F. Reinhard, J. Wrachtrup, and C.A. Meriles, “High-resolution correlation spectroscopy of 13C spins near a nitrogen-vacancy centre in diamond,” Nat Commun **4**, 1651 (2013).
[19] A. Laraoui, J.S. Hodges, C.A. Ryan, and C.A. Meriles, “Diamond nitrogen-vacancy center as a probe of random fluctuations in a nuclear spin ensemble,” Phys. Rev. B **84**(10), 104301 (2011).
[20] A. Laraoui, J.S. Hodges, and C.A. Meriles, “Magnetometry of random ac magnetic fields using a single nitrogen-vacancy center,” Appl. Phys. Lett. **97**(14), 143104 (2010).
[21] C.L. Degen, F. Reinhard, and P. Cappellaro, “Quantum sensing,” Rev. Mod. Phys. **89**(3), 035002 (2017).
[22] T. Gefen, A. Rotem, and A. Retzker, “Overcoming resolution limits with quantum sensing,” Nat Commun **10**, 4992 (2019).
[23] A. Cooper, W.K.C. Sun, J.-C. Jaskula, and P. Cappellaro, “Environment-assisted Quantum-enhanced Sensing with Electronic Spins in Diamond,” Phys. Rev. Applied **12**(4), 044047 (2019).
[24] C.-J. Yu, S. Von Kugelgen, D.W. Laorenza, and D.E. Freedman, “A Molecular Approach to Quantum Sensing,” ACS Cent. Sci. **7**(5), 712–723 (2021).
[25] F. Casola, T. van der Sar, and A. Yacoby, “Probing condensed matter physics with magnetometry based on nitrogen-vacancy centres in diamond,” Nat Rev Mater **3**, 1–13 (2018).
[26] A. Laraoui, and K. Ambal, “Opportunities for nitrogen-vacancy-assisted magnetometry to study magnetism in 2D van der Waals magnets,” Appl. Phys. Lett. **121**(6), 060502 (2022).
[27] A. Erickson, S.Q. Abbas Shah, A. Mahmood, I. Fescenko, R. Timalsina, C. Binek, and A. Laraoui, “Nanoscale imaging of antiferromagnetic domains in epitaxial films of $Cr_2O_3$ *via* scanning diamond magnetic probe microscopy,” RSC Adv. **13**(1), 178–185 (2023).
[28] R. Timalsina, H. Wang, B. Giri, A. Erickson, X. Xu, and A. Laraoui, “Mapping of Spin-Wave Transport in Thulium Iron Garnet Thin Films Using Diamond Quantum Microscopy,” Adv Elect Materials **10**(3), 2300648 (2024).
[29] A. Laraoui, H. Aycock-Rizzo, Y. Gao, X. Lu, E. Riedo, and C.A. Meriles, “Imaging thermal conductivity with nanoscale resolution using a scanning spin probe,” Nat Commun **6**, 8954 (2015).
[30] N. Aslam, H. Zhou, E.K. Urbach, M.J. Turner, R.L. Walsworth, M.D. Lukin, and H. Park, “Quantum sensors for biomedical applications,” Nat Rev Phys, 157-169 (2023).
[31] B.S. Miller, L. Bezinge, H.D. Gliddon, D. Huang, G. Dold, E.R. Gray, J. Heaney, P.J. Dobson, E. Nastouli, J.J.L. Morton, and R.A. McKendry, “Spin-enhanced nanodiamond biosensing for ultrasensitive diagnostics,” Nature **587**, 588–593 (2020).
[32] L. Nie, A.C. Nusantara, V.G. Damle, R. Sharmin, E.P.P. Evans, S.R. Hemelaar, K.J. Van Der Laan, R. Li, F.P. Perona Martinez, T. Vedelaar, M. Chipaux, and R. Schirhagl, “Quantum monitoring of cellular metabolic activities in single mitochondria,” Sci. Adv. **7**(21), eabf0573 (2021).

[33] Y. Wu, F. Jelezko, M.B. Plenio, and T. Weil, “Diamond Quantum Devices in Biology,” Angew Chem Int Ed **55**(23), 6586–6598 (2016).
[34] I. Lovchinsky, A.O. Sushkov, E. Urbach, N.P. De Leon, S. Choi, K. De Greve, R. Evans, R. Gertner, E. Bersin, C. Müller, L. McGuinness, F. Jelezko, R.L. Walsworth, H. Park, and M.D. Lukin, “Nuclear magnetic resonance detection and spectroscopy of single proteins using quantum logic,” Science **351**(6275), 836–841 (2016).
[35] C.L. Degen, M. Poggio, H.J. Mamin, C.T. Rettner, and D. Rugar, “Nanoscale magnetic resonance imaging,” Proc. Natl. Acad. Sci. U.S.A. **106**(5), 1313–1317 (2009).
[36] T. Staudacher, F. Shi, S. Pezzagna, J. Meijer, J. Du, C.A. Meriles, F. Reinhard, and J. Wrachtrup, “Nuclear Magnetic Resonance Spectroscopy on a (5-Nanometer) $^3$ Sample Volume,” Science **339**(6119), 561–563 (2013).
[37] P. Kehayias, A. Jarmola, N. Mosavian, I. Fescenko, F.M. Benito, A. Laraoui, J. Smits, L. Bougas, D. Budker, A. Neumann, S.R.J. Brueck, and V.M. Acosta, “Solution nuclear magnetic resonance spectroscopy on a nanostructured diamond chip,” Nat Commun **8**, 188 (2017).
[38] J. Smits, J.T. Damron, P. Kehayias, A.F. McDowell, N. Mosavian, I. Fescenko, N. Ristoff, A. Laraoui, A. Jarmola, and V.M. Acosta, “Two-dimensional nuclear magnetic resonance spectroscopy with a microfluidic diamond quantum sensor,” Sci. Adv. **5**(7), eaaw7895 (2019).
[39] I. Fescenko, A. Laraoui, J. Smits, N. Mosavian, P. Kehayias, J. Seto, L. Bougas, A. Jarmola, and V.M. Acosta, “Diamond Magnetic Microscopy of Malarial Hemozoin Nanocrystals,” Phys. Rev. Applied **11**(3), 034029 (2019).
[40] S. Lamichhane, K.A. McElveen, A. Erickson, I. Fescenko, S. Sun, R. Timalsina, Y. Guo, S.-H. Liou, R.Y. Lai, and A. Laraoui, “Nitrogen-vacancy magnetometry of individual Fe-triazole spin crossover nanorods,” (2023).
[41] S. Steinert, F. Ziem, L.T. Hall, A. Zappe, M. Schweikert, N. Götz, A. Aird, G. Balasubramanian, L. Hollenberg, and J. Wrachtrup, “Magnetic spin imaging under ambient conditions with sub-cellular resolution,” Nat Commun **4**, 1607 (2013).
[42] D.A. Simpson, R.G. Ryan, L.T. Hall, E. Panchenko, S.C. Drew, S. Petrou, P.S. Donnelly, P. Mulvaney, and L.C.L. Hollenberg, “Electron paramagnetic resonance microscopy using spins in diamond under ambient conditions,” Nat Commun **8**, 458 (2017).
[43] E. Schäfer-Nolte, L. Schlipf, M. Ternes, F. Reinhard, K. Kern, and J. Wrachtrup, “Tracking Temperature-Dependent Relaxation Times of Ferritin Nanomagnets with a Wideband Quantum Spectrometer,” Phys. Rev. Lett. **113**(21), 217204 (2014).
[44] P. Wang, S. Chen, M. Guo, S. Peng, M. Wang, M. Chen, W. Ma, R. Zhang, J. Su, X. Rong, F. Shi, T. Xu, and J. Du, “Nanoscale magnetic imaging of ferritins in a single cell,” Sci. Adv. **5**(4), eaau8038 (2019).
[45] S. Lamichhane, R. Timalsina, C. Schultz, I. Fescenko, K. Ambal, S.-H. Liou, R.Y. Lai, and A. Laraoui, “Nitrogen-Vacancy Magnetic Relaxometry of Nanoclustered Cytochrome C Proteins,” Nano Lett. 24(3), 873-880 (2024).
[46] F. Perona Martínez, A.C. Nusantara, M. Chipaux, S.K. Padamati, and R. Schirhagl, “Nanodiamond Relaxometry-Based Detection of Free-Radical Species When Produced in Chemical Reactions in Biologically Relevant Conditions,” ACS Sens **5**(12), 3862–3869 (2020).
[47] A. Ermakova, G. Pramanik, J.-M. Cai, G. Algara-Siller, U. Kaiser, T. Weil, Y.-K. Tzeng, H.C. Chang, L.P. McGuinness, M.B. Plenio, B. Naydenov, and F. Jelezko, “Detection of a Few Metallo-Protein Molecules Using Color Centers in Nanodiamonds,” Nano Lett. **13**(7), 3305–3309 (2013).
[48] F. Gorrini, R. Giri, C.E. Avalos, S. Tambalo, S. Mannucci, L. Basso, N. Bazzanella, C. Dorigoni, M. Cazzanelli, P. Marzola, A. Miotello, and A. Bifone, “Fast and Sensitive Detection of Paramagnetic Species Using Coupled Charge and Spin Dynamics in Strongly Fluorescent Nanodiamonds,” ACS Appl. Mater. Interfaces **11**(27), 24412–24422 (2019).
[49] P. Hensley, S.J. Edelstein, D.C. Wharton, and Q.H. Gibson, “Conformation and spin state in methemoglobin,” J Biol Chem **250**(3), 952–960 (1975).
[50] F. Mehraban, S. Rayati, V. Mirzaaghaei, and A. Seyedarabi, “Highlighting the Importance of Water Alkalinity Using Phosphate Buffer Diluted With Deionized, Double Distilled and Tap Water, in Lowering

Oxidation Effects on Human Hemoglobin Ozonated at High Ozone Concentrations in vitro," Front. Mol. Biosci. **7**, 543960 (2020).
[51] M.R. Dayer, M.S. Dayer, and A.A. Moosavi- Movahedi, "A tri state mechanism for oxygen release in fish hemoglobin: Using Barbus sharpeyi as a model," Mol Biol Res Commun **3**(2), 101–113 (2014).
[52] Kamaljeet, S. Bansal, and U. SenGupta, "A Study of the Interaction of Bovine Hemoglobin with Synthetic Dyes Using Spectroscopic Techniques and Molecular Docking," Front. Chem. **4**, (2017).
[53] Y.-X. Huang, Z.-J. Wu, B.-T. Huang, and M. Luo, "Pathway and Mechanism of pH Dependent Human Hemoglobin Tetramer-Dimer-Monomer Dissociations," PLoS ONE **8**(11), e81708 (2013).
[54] H.P. Erickson, "Size and Shape of Protein Molecules at the Nanometer Level Determined by Sedimentation, Gel Filtration, and Electron Microscopy," Biol Proced Online **11**, 32–51 (2009).
[55] A. Sato-Tomita, and N. Shibayama, "Size and Shape Controlled Crystallization of Hemoglobin for Advanced Crystallography," Crystals **7**(9), 282 (2017).
[56] F.C. Ziem, N.S. Götz, A. Zappe, S. Steinert, and J. Wrachtrup, "Highly Sensitive Detection of Physiological Spins in a Microfluidic Device," Nano Lett. **13**(9), 4093–4098 (2013).
[57] E.E. Kleinsasser, M.M. Stanfield, J.K.Q. Banks, Z. Zhu, W.-D. Li, V.M. Acosta, H. Watanabe, K.M. Itoh, and K.-M.C. Fu, "High density nitrogen-vacancy sensing surface created via He+ ion implantation of 12C diamond," Appl. Phys. Lett. **108**(20), 202401 (2016).
[58] A. Jarmola, V.M. Acosta, K. Jensen, S. Chemerisov, and D. Budker, "Temperature- and Magnetic-Field-Dependent Longitudinal Spin Relaxation in Nitrogen-Vacancy Ensembles in Diamond," Phys. Rev. Lett. **108**(19), 197601 (2012).
[59] M. Mrózek, D. Rudnicki, P. Kehayias, A. Jarmola, D. Budker, and W. Gawlik, "Longitudinal spin relaxation in nitrogen-vacancy ensembles in diamond," EPJ Quantum Technol. **2**, 22 (2015).
[60] A. Laraoui, J.S. Hodges, and C.A. Meriles, "Nitrogen-Vacancy-Assisted Magnetometry of Paramagnetic Centers in an Individual Diamond Nanocrystal," Nano Lett. **12**(7), 3477–3482 (2012).
[61] A. Laraoui, and C.A. Meriles, "Approach to Dark Spin Cooling in a Diamond Nanocrystal," ACS Nano **7**(4), 3403–3410 (2013).
[62] F.A. Freire-Moschovitis, R. Rizzato, A. Pershin, M.R. Schepp, R.D. Allert, L.M. Todenhagen, M.S. Brandt, A. Gali, and D.B. Bucher, "The Role of Electrolytes in the Relaxation of Near-Surface Spin Defects in Diamond," ACS Nano **17**(11), 10474–10485 (2023).
[63] K. Victor, A. Van-Quynh, and R.G. Bryant, "High Frequency Dynamics in Hemoglobin Measured by Magnetic Relaxation Dispersion," Biophys J **88**(1), 443–454 (2005).
[64] M. Xie, X. Yu, L.V.H. Rodgers, D. Xu, I. Chi-Durán, A. Toros, N. Quack, N.P. De Leon, and P.C. Maurer, "Biocompatible surface functionalization architecture for a diamond quantum sensor," Proc. Natl. Acad. Sci. U.S.A. **119**(8), e2114186119 (2022).
[65] R.D. Allert, F. Bruckmaier, N.R. Neuling, F.A. Freire-Moschovitis, K.S. Liu, C. Schrepel, P. Schätzle, P. Knittel, M. Hermans, and D.B. Bucher, "Microfluidic quantum sensing platform for lab-on-a-chip applications," Lab Chip **22**(24), 4831–4840 (2022).
[66] M. Fujiwara, "Diamond quantum sensors in microfluidics technology," Biomicrofluidics **17**(5), 054107 (2023).
[67] S. Sangtawesin, B.L. Dwyer, S. Srinivasan, J.J. Allred, L.V.H. Rodgers, K. De Greve, A. Stacey, N. Dontschuk, K.M. O'Donnell, D. Hu, D.A. Evans, C. Jaye, D.A. Fischer, M.L. Markham, D.J. Twitchen, H. Park, M.D. Lukin, and N.P. de Leon, "Origins of Diamond Surface Noise Probed by Correlating Single-Spin Measurements with Surface Spectroscopy," Phys. Rev. X **9**(3), 031052 (2019).